\documentstyle[12pt]{article}
\newcommand{\Ebf}{\mbox{\boldmath $E$}}

\newcommand{\Hbf}{\mbox{\boldmath $H$}}

\newcommand{\zbf}{\mbox{\boldmath $z$}}

\newcommand{\nabf}{\mbox{\boldmath $\nabla$}}
\newcommand{\rhobf}{\mbox{\boldmath $\rho$}}
\newcommand{\TM}{{\rm TM}}

\newcommand{\ph}{{\rm ph}}
\newcommand{\Drm}{{\rm D}}
\newcommand{\orm}{{\rm o}}
\newcommand{\pa}{\partial}

\newcommand{\X}{{\rm X}}
\newcommand{\pisr}{{\sqrt{\pi}}}

\newcommand{\text}{\rm}

\newcommand{\drm}{{\rm d}}
\newcommand{\grm}{{\rm g}}

\newcommand{\et}{{\rm and}}

\newcommand{\ug}{ \; = \; }
\newcommand{\ugg}{ \ = \ }

\newcommand{\infi}{\infty}

\newcommand{\lra}{\longrightarrow}

\newcommand{\bb}{\begin{equation}}
\newcommand{\ee}{\end{equation}}
\newcommand{\bega}{\begin{eqnarray}}
\newcommand{\ega}{\end{eqnarray}}
\newcommand{\begae}{\begin{eqnarray*}}
\newcommand{\egae}{\end{eqnarray*}}

\newcommand{\h}{\hspace*{4ex}}
\newcommand{\dis}{\displaystyle}

\newcommand{\be}{\beta}

\newcommand{\th}{\theta}

\newcommand{\om}{\omega}

\newcommand{\cent}{\centerline}
\newcommand{\vs}{\vspace*}

\begin{document}

\baselineskip 0.65cm

\begin{center}

{\large {\bf Superluminal X-shaped beams propagating without
distortion along a coaxial guide}$^{\: (\dag)}$}
\footnotetext{$^{\: (\dag)}$
%%% Electronic version in the Lanl Archives
%%% \# physics/*******. \
Work partially supported by FAPESP (Brasil) and by INFN, MIUR
(Italy). \ E-mail addresses for contacts: recami@mi.infn.it [ER];
giz.r@uol.com.br [MZR].}

\end{center}

\vs{2mm}

\cent{ Michel Zamboni-Rached, \ K. Z. N\'obrega}

\vs{0.2 cm}

\centerline{{\em DMO--FEEC, State University at Campinas,
Campinas, S.P., Brasil.}}

\vs{0.3 cm}

\cent{ Erasmo Recami }

\vs{0.2 cm}

\cent{{\em Facolt\`a di Ingegneria, Universit\`a statale di Bergamo,
Dalmine (BG), Italy;}}
\cent{{\em INFN---Sezione di Milano, Milan, Italy; \ {\rm and}}}
\cent{{\em C.C.S., State University at Campinas,
Campinas, S.P., Brasil.}}

\vs{0.2 cm}

\centerline{\rm and}

\vs{0.2 cm}

\cent{ Hugo Enrique Hern\'{a}ndez-Figueroa }

\vs{0.2 cm}

\cent{{\em DMO--FEEC, State University at Campinas,
Campinas, S.P., Brasil.}}

\vs{0.1 cm}

\

{\bf Abstract  \ --} \ In a previous paper we showed that
localized Superluminal solutions to the Maxwell equations
exist, which propagate down (non-evanescence) regions of a
metallic cylindrical waveguide. In this paper we construct analogous
non-dispersive waves propagating along coaxial cables. Such
new solutions, in general, consist in trains of (undistorted)
Superluminal ``X-shaped" pulses. Particular attention is paid
to the construction of finite total energy solutions. Any
results of this kind may find application in the other fields
in which an essential role is played by a wave-equation (like
acoustics, geophysics, etc.).\\



\newpage

{\bf 1. -- Introduction}\\

\h In a previous paper[1] we constructed localized Superluminal solutions to
the Maxwell equations propagating along (non-evanescent regions of) a
metallic
cylindrical waveguide. In the present paper we are going to show that
analogous solutions exist even for metallic {\em coaxial} cables. Their
interest is due to the fact that they propagate without distortion with
Superluminal group-velocity.

\h Let us recall that already in 1915 Bateman[2] showed
Maxwell equations to admit (besides of the ordinary solutions,
endowed with speed $c$ in vacuum) of wavelet-type solutions,
endowed in vacuum with group-velocities $0 \leq v \leq c$. \
But Bateman's work went practically unnoticed, with the exception
of a few authors as Barut et al.[3]. (Incidentally, Barut et al.
even constructed a wavelet-type solution[4] traveling with
Superluminal group-velocity $V > c$).

\h In recent times, however, many authors started to discuss
the circumstance that all  wave equations admit
of solutions with $0 \leq v \leq \infi$: see, e.g., refs.[5]. \
Most of those authors confined themselves to investigate
(sub- or Super-luminal) non-dispersive solutions propagating in the
open space only: namely, those solutions that
had been called ``undistorted progressive waves" by Courant \&
Hilbert[6].

\h Among localized solutions, the most interesting appeared
to be the {\em ``$\X$-shaped waves"}, which, predicted long ago
to exist within Special Relativity in its extended version[7,8],
had been mathematically constructed by Lu et al.[9] for
acoustic waves, and by Ziolkowski et al.[10] and by Recami[11]
for electromagnetic waves.

\h Let us stress that such ``X-shaped" localized solutions
are Superluminal (i.e., travel with a speed larger than $c$
in vacuum) in the electromagnetic case; and are ``Super-sonic"
(i.e., travel with a speed larger than the sound-speed in the
medium) in the acoustic case. \ The first authors to produce
{\em experimentally} X-shaped waves were Lu et al.[13] for
acoustics, Saari et al.[14] for optics, and Mugnai et al.[15]
for microwaves. \ Let us also emphasize, incidentally, that
all such solutions can have an interesting role even in
seismology, and probably in the gravitational wave sector.

\h Notwithstanding all that work[16], it is not well understood yet what
solutions ---let us now confine ourselves, for simplicity, to Maxwell
equations and to electromagnetic waves only--- have to enter into the play
in realistic experiments using waveguides, optical fibers, etc.\\

\

{\bf 2. -- The case of a cylindrical waveguide}\\

\h As we already mentioned, in ref.[1] we constructed, for the TM (transverse
magnetic) case, localized solutions to the Maxwell equations which propagate
(undistorted) with Superluminal speed along a cylindrical waveguide. Let
us take advantage of the present opportunity for calling further attention to
two points, which received just a mention in ref.[1], with regard to eq.(9)
and Fig.2 therein. Namely, let us here stress that:

\h (i) those solutions
consist in a {\em train} of pulses like the one depicted in Fig.2 of
ref.[1];  and that

\h (ii) each of such pulses is {\em $\X$-shaped}.

A more complete representation of the TM (and TE) non-dispersive waves,
traveling down a cylindrical waveguide, will be forwarded elsewhere.\\

\

{\bf 3. -- The case of a coaxial cable}\\

\h Let us now examine the case of a coaxial cable (a metallic coaxial
waveguide, to fix our ideas), that is, of the region delimited by two
cylinders with radius $\rho=r_1$ and $\rho=r_2$, respectively, and axially
symmetric with respect to the $z$-axis: see Fig.1. \ We shall consider in
this article both the TM case, characterized by the Dirichlet boundary
conditions[17] (for any time instant $t$)

\

\hfill{$
E_z(\rho=r_1; \, t) \ug 0 \ ; \ \ \ \ \ \ \ \  E_z(\rho=r_2; \, t) \ug 0 \ ;
$\hfill} (1)

\

and the TE (transverse electric) case, characterized by the Neumann boundary
conditions[17] (for any $t$)

\

\hfill{$
\dis{{\pa \over {\pa\rho}} H_z(\rho=r_1; \, t)} \ug 0 \ ; \ \ \ \ \ \ \ \
\dis{{\pa \over {\pa\rho}} H_z(\rho=r_2; \, t)} \ug 0 \ .
$\hfill} (2)

\

\h To such aims, we shall first generalize a theorem due to Lu et al.[18]
(stated and demonstrated below, in the Appendix),
which showed how to start from a solution holding {\em in the plane}
($x,y$) for constructing a three-dimensional solution rigidly moving
along the $z$-axis with Superluminal speed $V$.  The Lu et al.'s theorem
was valid for the vacuum. In ref.[1] we set forth its generalization for
a cylindrical waveguide, while here we are going to extend it, as we
said above, for a coaxial cable. \ Let us first recall what Lu et al.'s
theorem is about. \ If we assume that $\psi(\rhobf;t)$, with $\rhobf \equiv
(x,y)$, is a solution of the two-dimensional homogeneous wave equation

\

\hfill{$
\left( \pa_x^2 + \pa_y^2 - \dis{{1 \over {c^2}}} \pa_t^2 \right) \;
\psi(\rhobf;t) \ = \ 0 \ ,
$\hfill} (3)

\

then, by applying the transformations

\

\hfill{$
\rhobf \lra \rhobf \, \sin \th \ ;  \ \ \ \et \ \ \ t \lra t -
z \, \dis{({{\cos \th} \over c})} \ ,
$\hfill} (4)

\

the angle $\th$ being fixed, with $0 < \th < \pi/2$, one gets[18] that
$\psi(\rhobf \sin \th; \, t-z \cos \th / c)$ is now a solution of the
{\em three-dimensional} homogeneous wave-equation

\

\hfill{$
\left( \nabf^2 - {1 \over {c^2}} \pa_t^2 \right) \ \psi\!\left( \rhobf  \;
\sin\th ; \ t - z \, {\dis{\cos\th} \over  c} \right) = 0 \ ,
$\hfill} (5)

\

where now $\nabf^2 \equiv \pa_x^2 + \pa_y^2 + \pa_z^2$; \ $\rhobf \equiv
(x,y)$.

\h The mentioned theorem holds for the free case, so that in general it
does not hold when introducing boundary conditions. We shall see, however,
that it can be extended even to the case of a two-dimensional solution
$\psi$ valid on an annular domain, $a \leq \rho \leq b; \ \rho \equiv
|\rhobf|$, with either the (Dirichlet) boundary conditions

\

\hfill{$
\psi(\rho=a; \, t) \ug \psi(\rho=b; \, t) \ug 0 \ ,
$\hfill} (1')

\

or the (Neumann) boundary conditions

\

\hfill{$
\dis{{\pa \over {\pa\rho}} \psi(\rho=a; \, t)} \ug
\dis{{\pa \over {\pa\rho}} \psi(\rho=b; \, t)} \ug 0 \ .
$\hfill} (2')

\

\h Let us notice right now that transformations (4), with condition (1') or
(2'), lead to a (three-dimensional) solution rigidly traveling with
Superluminal speed $V= c / \cos \th$ inside a coaxial cable with internal
and external radius equal (no longer to $a$, $b$, but) to \ $r_1 = a / \sin \th >
a$ \ and \ $r_2 = b / \sin \th > b$, \ respectively. \ The same procedure can
be applied also in other cases, provided that the boundary conditions depend
on $x, \ y$ only: as in the case, e.g., of a cable with many cylindrical
(empty) tunnels inside it.\\

\

{\bf 4. -- The transverse magnetic (TM) case}\\

\h Let us go back to the two-dimensional equation (3) with the boundary
conditions (1'). Let us choose for instance the simple {\em initial}
conditions $\psi(\rho; \, t=0) \equiv \phi(\rho)$ and $\pa \psi / \pa t
\equiv \xi(\rho)$ at $t=0$, where

\

\hfill{$
\phi(\rho) \ug \delta (\rho - \rho_0) \ ; \ \ \ \ \ \
\dis{\left. \xi(\rho) \right|_{t=0 }} \ug 0
$\hfill} (6)

\

with

\

\hfill{$
a < \rho_0 < b \ .
$\hfill} (6')

\

\h Following a method similar to the one in ref.[1], and using
the boundary conditions (1'), in cylindrical co-ordinates and for
axial symmetry one gets solutions to eq.(3) of the type $\psi =
\sum R_n(\rho) \; T_n(t)$ in the following form:

\

\hfill{$
2 \, \psi(\rho;t) \ugg \sum_{n=1}^{\infty} R_n(\rho) \; [A_n \, \cos \om_n t
- B_n \, \sin \om_n t] \ ,
$\hfill} (7)

\

where the functions $R(\rho)$ are

\

\hfill{$
R_n(\rho) \; \equiv \; N_0(k_n a) \; J_0(k_n \rho) - J_0(k_n a) \; N_0(k_n
\rho) \ ,
$\hfill} (8)

\

quantities $N_0$ and $J_0$ being the zeroth-order Neumann and
Bessel functions, respectively; \ and where the characteristic
angular frequencies[19] can be evaluated numerically, they being
solutions to the equation [$\om_n = c k_n$]

\

\hfill{$
\dis{{{J_0(k_n a)} \over {N_0(k_n a)}} \ = \ {{J_0(k_n b)} \over {N_0(k_n
b)}} } \ .
$\hfill} (9)

\

\h The initial conditions (6) imply that $\sum A_n \, R_n(\rho) =
\delta (\rho - \rho_0)$, and $\sum B_n \, R_n(\rho) = 0$, so that all the
coefficients $B_n$ vanish, and eventually one obtains the two-dimensional
solution

\

\hfill{$ \Psi_{2\Drm}(\rho;t) \ugg \sum_{n=1}^{\infty} A_n \;
R_n(\rho) \; \cos \om_n t \ , $\hfill} (10)

\

with

\

\hfill{$
\begin{array}{l}
2 \, A_n \ = \ \left\{- a^2 \, \left[ N_0(k_n a) J_1(k_n a) -  J_0(k_n a)
N_1(k_n a)
\right]^2 \; + \right. \\
\;\;\;\;\;\;\;\;\;\;\;\;\;\;\;+ \; b^2 \;  \left. \left[ N_0(k_n
a) J_1(k_n b) - J_0(k_n a) N_1(k_n b) \right]^2  \right\}^{-1}
R_n(\rho_0) \ .
\end{array}
$\hfill} (11)

\

\h One can notice that the present procedure is mathematically analogous to
the analysis of the free vibrations of a ring-shaped elastic membrane[19].

\h For any practical purpose, one has of course to take a finite number $N$
of terms in expansion (10). In Fig.2 we show, e.g., the two-dimensional
functions $|\Psi_{2\Drm}(\rho;t)|^2$ of eq.(10) for fixed time $(t=0)$ and
for $N=10$ (dotted line) or $N=40$ (solid line). \ Notice that when the value
$N$ is finite, the first one of conditions (6) is no longer a delta function,
but represents a physical wave, which nevertheless is still clearly bumped
(Fig.2). \ It is rather interesting that, for each value of $N$, one meets a
different (physical) situation; at the extent that we obtain infinite
many {\em different} families of three-dimensional solutions, by varying the
truncating value $N$ in eq.(12) below.

\h Actually, by the transformations (4) we arrive from eq.(10) at the
three-dimensional Superluminal non-dispersive solution $\Psi_{3\Drm}$,
propagating without distortion along a metallic coaxial waveguide, i.e.,
down a coaxial cable [$V > c$]:

\

\hfill{$
\Psi_{3\Drm}(\rho; \, z-Vt) \ug \sum_{n=1}^{\infty} A_n \; R_n(\rho\sin\th)
\; \cos \left[ k_n \, (z-Vt) \, \cos \th \right]
$\hfill} (12)

\

which is a sum over different propagating modes. The fact that $V
= c / \cos \th > c$ means (once more) that the
group-velocity\footnote{Let us recall that the group-velocity is
well defined only when the pulse has a clear bump in space; but
it can be calculated by the approximate relation $v_g \simeq \drm
\om / \drm \be$, quantity $\be$ being the wavenumber, {\em only}
when some extra conditions are satisfied (namely, when $\om$ as a
function of $\be$ is also clearly bumped). In the present case the
group-velocity is very well defined, but cannot be evaluated
through that simple relation, since $\om$ is a discrete function
of $\be$: cf. eq.(9) and Sect.6, eq.(22), below.} of our pulses
is Superluminal. \ For simplicity, in our Figures we shall put $z
- Vt \equiv \eta$.

\h Let us notice that transformations (4), which change ---as we already
know--- $a$ into $r_1=a / \sin \th$ and $b$ into $r_2=b / \sin \th$, are
such that the maximum of $\Psi_{3\Drm}$ is got for the value $\rho_0 /
\sin \th$ of $\rho$. However, solution (12) does automatically satisfy
on the cylinders with radius $r_1$ and $r_2$ the conditions [$\Psi_{3\Drm}
\equiv E_z$]:

$$
\Psi_{3\Drm}(\rho=a/\sin\th, \, z; \, t) \ug \Psi_{3\Drm}(\rho=b/\sin\th, \,
z; \, t) \ug 0 \ .$$

\h Till now, $\Psi_{3\Drm}$ has represented the electric field component
$E_z$. \ Let us add that in the TM case[20]:

\

\hfill{$
\Ebf_\bot \ug i \, \dis{ {c\,V \over {V^2-c^2 }} \; \sum_{n=1}^{\infty} {1
\over k_n} \,
\nabf_\bot \Psi_{3\Drm} } \ ,
$\hfill} (12a)

\

where

$$\dis{ {cV \over {V^2-c^2 }} \; \equiv \; {{\cos \th} \over {\sin^2 \th}} }
\ , \;\;\;\; k_n=\om_n/c \,\, , $$

and

\

\hfill{$
\Hbf_\bot \ug \dis{  \varepsilon_0\, \frac{V}{c} } \; {{\hat{\zbf}} \wedge
\Ebf_\bot}
$\hfill} (12b)

\

\h As we mentioned above, for any truncating value $N$ in expansion (10), we
get a {\em different} physical situation: In a sense, we excite in a
{\em different} way the two-dimensional annular membrane, obtaining (via
Lu et al.'s theorem) different three-dimensional solutions, which
correspond[1] to nothing but summation (12) truncated at the value $N$.

\h In Figs.3a,b, we show a single (X-shaped) three-dimensional pulse
$\Psi_{3\Drm}$ with $\th = 84^\orm$, and $N=10$  or $N=40$, respectively.

\h In Fig.4, by contrast, we depict a couple of elements of the {\em train}
of X-shaped pulses represented by eq.(12), for $\th = 45^\orm$ and $N=40$.

\h In Fig.5 the orthogonal projection is moreover shown of a
single pulse (of the solution in Fig.4) onto the ($\rho,z$) plane
for $t=0$, with $\th = 45^\orm$ and $N=40$. \ Quantities $\rho$ and $\eta$
are always in centimeters.\\

\

{\bf 5. -- The transverse electric (TE) case}\\

\h In the TE case, one has to consider the two-dimensional equation (3) with
the boundary conditions (2'), while the initial conditions (6) can remain
the same.

\h As in Sect.4, one gets ---still for axial symmetry in cylindrical
co-ordinates--- the following solution to eq.(3):

\

\hfill{$
2 \, \psi(\rho;t) \ugg \sum_{n=1}^{\infty} R_n(\rho) \; [A_n \, \cos \om_n t
- B_n \, \sin \om_n t] \ ,
$\hfill} (13)

\

where now the functions $R_n(\rho)$ are

\

\hfill{$ R_n(\rho) \equiv N_1(k_n a) \, J_0(k_n \rho) - J_1(k_n
a) \, N_0(k_n \rho) \ , $\hfill} (13')

\

defined in terms of different values of $k_n$. In fact, the
characteristic (angular) frequencies are now to be obtained by
the new relation

\

\hfill{$
\dis{{{J_1(k_n a)} \over {N_1(k_n a)}} \ = \ {{J_1(k_n b)} \over {N_1(k_n
b)}} } \ .
$\hfill} (14)

\

\h Again, the initial conditions (6) entail that $\sum A_n \, R_n(\rho) =
\delta (\rho - \rho_0)$, and $\sum B_n \, R_n(\rho) = 0$, so that all the
coefficients $B_n$ vanish, and one gets the two-dimensional solution

\

\hfill{$ \Psi_{2\Drm}(\rho;t) \ugg \sum_{n=1}^{\infty} A_n \;
R_n(\rho) \; \cos \omega_n t \ , $\hfill} (15)

\

where the coefficients $A_n$ are given by

\

\hfill{$
\begin{array}{l}
2 \, A_n \ = \ \left\{- a^2 \, \left[ N_1(k_n a) J_0(k_n a) -
J_1(k_n a) N_0(k_n a) \right]^2 \; + \right. \\
\;\;\;\;\;\;\;\;\;\;\;\;\;\;\;+ \; b^2 \;  \left. \left[ N_1(k_n
a) J_0(k_n b) - J_1(k_n a) N_0(k_n b) \right]^2  \right\}^{-1}
R_n(\rho_0) \ .
\end{array}
$\hfill} (15')

\

\h In this case one obtains, by transformations (4), the Superluminal
non-dispersive three-dimensional solution

\

\hfill{$
\Psi_{3\Drm}(\rho; \, z-Vt) \ug \sum_{n=1}^{\infty} A_n \; R_n(\rho\sin\th) \;
\cos \left[ k_n \, (z-Vt) \, \cos \th \right]
$\hfill} (16)

\

propagating along the metallic coaxial waveguide with group-velocity
$V=c/\cos\th>c$. The present solution (16) satisfies the boundary conditions

$$\dis{ \left. {\pa \over {\pa\rho}} \Psi_{3\Drm}(\rho,z,t) \right|_{\rho=
{a\over{\sin\th}}}} \ugg \dis{ \left. {\pa \over {\pa\rho}}
\Psi_{3\Drm}(\rho,z,t) \right|_{\rho={b\over{\sin\th}}}} \ugg 0 \ ,$$

where now $\Psi_{3\Drm} \equiv H_z$. \ The transverse components, in the TE
case, are given[20] by

\

\hfill{$
\Hbf_\bot \ug  \dis{ {-c\,V \over {V^2-c^2 }} \;
\sum_{n=1}^{\infty} {1 \over k_n} \,\sin \left[ k_n \, (z-Vt) \,
\cos \th \right]\, \nabf_\bot R_n(\rho\sin\th) } \ ,
$\hfill} (17a)

\

and

\

\hfill{$
\Ebf_\bot \ug \dis{ - \mu_0\, \frac{V}{c}} \; {{\hat{\zbf}} \wedge
\Hbf_\bot}
$\hfill} (17b)

\

\h In Fig.6 we plot our function $\Psi_{2\Drm}$ with $N=10$ (dotted line)
or $N=40$ (solid line). \ In Fig.7 there are depicted, by contrast,
two elements of the train of X-shaped pulses represented by eq.(16),
with $\th = 60^\orm$, for $N=40$ only. \ In Fig.8, at last, we
show the orthogonal projection (of a single pulse of the solution in Fig.7)
onto the plane ($\rho,z$) for $t=0$, with $\th = 60^\orm$ and
$N=40$. \ Quantities $\rho$ and $\eta$ are in cm.\\

\

{\bf 6. -- Rederivation of our results from the standard theory of waveguide
propagation}\\

\h Lu's theorem is certainly a very useful tool to build up localized
solutions to Maxwell equations: nevertheless, due to the novelty of our
previous results, it may be worthwhile to outline an alternative
derivation[1] of them which can sound more familiar. To such an aim, we shall
follow the procedure introduced in ref.[1].

\h For the sake of simplicity, let us limit ourselves to the domain of TM
(transverse magnetic) modes. \ When a solution in terms of the longitudinal
electric component, $E_z$, is sought, one has to deal with the boundary
condition $E_z = 0$; we shall look, moreover, for axially symmetric solutions
(i.e., independent of the azimuth variable, $\varphi$): Such choices could be
easily generalized, just at the cost of increasing the mathematical
complexity. \ Quantity $E_z$ is then completely equivalent to the scalar
variable $\Psi \equiv \Psi_{3\Drm}$ used in the previous analysis.

\h Let us look for solutions of the form[1]

\

\hfill{$
E_z(\rho,z;t) \ugg K \; R(\rho) \; \exp \left[ i \left( \dis{{{\om z \cos\th}
\over c}} - \om t \right) \right]
$\hfill} (18)

\

where $R(\rho)$ is assumed to be a function of the radial coordinate $\rho$
only, and $K$ is a normalization constant. Here we call $c$ the velocity of
light in the medium filling the coaxial waveguide, supposing it
nondispersive. The (angular) frequency $\om$ is for the moment arbitrary.

\h By inserting expression (18) into the Maxwell equation for $E_z$, one
obtains[1]

\

\hfill{$
\rho^2 {{\drm^2 R(\rho)} \over {\drm \rho^2}} + \rho {{\drm R(\rho)} \over
{\drm \rho}} + \rho^2 \; \Omega^2 \; R(\rho) \ugg 0 \, ; \ \ \ \ \  \Omega
\equiv {{\om \sin \th} \over c} \; ,
$\hfill} (19)

\

whose only solution, which is finite on the waveguide axis, is \ $R(\rho) =
N_0(\om a / c) \; J_0(\om \rho \sin\th / c) - J_0(\om a / c) \;
N_0(\om \rho \sin\th / c)$, \ which is analogous to eq.(8).

\h By imposing the boundary conditions $R(\rho) = 0$ for $\rho =
r_1 = a / \sin\th$ and $\rho = r_2 = b / \sin\th$ , one gets the acceptable
frequencies from the characteristic equation:

\

\hfill{$ \dis{{{J_0(\om_n a /c)} \over {N_0(\om_n a
/c)}} \ = \ {{J_0(\om_n b /c)} \over {N_0(\om_n b
/c)}} } \ , $\hfill} (20)

\

so that one has a different function $R_n(\rho)$ for each value of $\om_n$. \
Therefore, assuming[1] an arbitrary parameter $\th$, we find that, for every
mode supported by the waveguide and labeled by the index $n$, there is just
one frequency at which the assumed dependence (18) on $z$ and $t$ is physically
realizable. \ Let us show such a solution to be the standard one known from
classical electrodynamics. In fact, by inserting[1] the allowed frequencies
$\om_n$ into the complete expression of the mode, we have:

\

\hfill{$
E_z^n(\rho,z;t) \ugg K \; R_n(\rho) \; \exp \left[ i
\left( \dis{{{\om_n z \cos\th} \over c}} - \om_n t \right) \right] \ .
$\hfill} (21)

\

\h But the generic solution for (axially symmetric) $\TM_{0n}$
modes[21] in a coaxial metallic waveguide is [$\Omega_n \equiv
\om_n \, \sin\th/c$]:

\

\hfill{$
E_z^{\TM_{0n}} \ugg K \; R_n(\rho) \; \exp \left[ i
\left( \be(\om_n) \, z - \om_n t \right) \right] \ ,
$\hfill} (22)

\

the wavenumber $\be$ being a discrete function of $\om$, with the
``dispersion relations"

$$\be^2(\om_n) \ug \frac{\om_n^2}{c^2} - \Omega_n^2 \ .$$.

By identifying \ $\be(\om_n) \equiv \om_n \, \cos\th/c$, \ as
suggested by eq.(21), and remembering the expression for $\om_n$
given by eq.(20), the ordinary dispersion relation is got[1]. We
have therefore verified that every term in the expansion (12) is
a solution to Maxwell equations not different from the usual one.

\h The uncommon feature of our solution (12) is that, given a particular
value of $\th$, the phase-velocity of {\em all} its terms is always the same, it
being independent of the mode index~$n$:

$$V_\ph \ug \left[ {{\be (\om_n)} \over {\om_n}} \right]^{-1} \ug {c \over
{\cos\th}} \ .$$

In such a case it is well-known that the group-velocity of the pulse
{\em equals} the phase-velocity[22]: and in our case is the velocity
{\em tout court} of the localized pulse.

\h With reference to Fig.9, we can easily see[1] that all the allowed values
of $\om_n$ can be calculated by determining the intersections of the various
branches of the dispersion relation with a straight line, whose slope
depends on $\th$ only. By using suitable combinations of terms, corresponding
to different indices $n$, as in our eq.(12), it is possible to describe a
disturbance having a time-varying profile[1], as already shown in Figs.3-4
above. Each pulse thus displaces itself {\em rigidly}, with a {\em velocity}
$v \equiv v_\grm$ equal to $V_\ph$.

\h It should be repeated that the velocity $v$ (or
group-velocity $v_\grm \equiv v$) of the pulses corresponding to eq.(9) is
not to be evaluated by the ordinary formula $v_\grm \simeq
\drm \om / \drm \be$ (valid for quasi-monochromatic signals). \ This is at
variance with the common situation in optical and microwave communications,
when the signal is usually an ``envelope" superimposed to a carrier wave
whose frequency is generally much higher than the signal bandwidth. In
{\em that} case the standard formula for $v_\grm$ yields the correct velocity
to deal with (e.g., when propagation delays are studied). \ Our case, on the
contrary, is much more reminiscent of a baseband modulated signal, as those
studied in ultrasonics: the very concept of a carrier becomes meaningless
here, as the discrete ``harmonic" components have widely different
frequencies[1].

\h Let us finally remark[1] that similar considerations could be extended to
all the situations where a waveguide supports several modes. Tests at microwave
frequencies should be rather easy to perform; by contrast, experiments in the
optical domain would face the problem of the limited extension of the
spectral windows corresponding to not too large attenuation, even if work[23]
is in progress in many directions.

\h Moreover, results of the kind presented in this paper, as well as in
refs.[1,11,12], may find application in the other fields in which an essential
role is played by a wave-equation (like acoustics, seismology, geophysics,
and relativistic quantum mechanics, possibly.).\\

\

{\bf 7. -- How to get finite total energy solutions}\\

\h We shall go on following the standard formalism of Sect.6;
what we are going to do holds, however, for both the TM and the
TE case. \ Let us anticipate that, in order to get finite total
energy solutions (FTES), we shall have to replace each
characteristic frequency $\om_n$ [cf. eq.(9), or eq.(14) or
rather Fig.9] by a {\em small} frequency band $\Delta \om$
centered at $\om_n$, always choosing the same $\Delta \om$
independently of $n$. In fact, since all the modes entering the
Fourier-type expansion (12), or (16), possess the same
phase-velocity $V_\ph \equiv V = c/\cos\th$, each small bandwidth
packet associated with $\om_n$ will possess the same
group-velocity $v_\grm = c^2/V_\ph$, so that we shall have as a
result a wave whose {\em envelope} travels with the {\em subluminal}
group-velocity $v_\grm$.  {\em However}, inside the subluminal
envelope, one or more {\em pulses} will be travelling with the
dual (Superluminal) speed $V=c^2/v_\grm $. Such well-localized
peaks have nothing to do with the ordinary (sinusoidal)
carrier-wave, and will be regarded as constituting {\em the
relevant} wave. \ Before going on, let us mention that previous
work related to FTESs can be found ---as far as we know--- only
in refs.[24] and [12].

\h Formally, to get FTESs, let us consider the ordinary (three-dimensional)
solutions for a coaxial cable:

\

\hfill{$
\psi_{n}(\rho,z;t) \ug K_n \; R_n(\rho) \; \cos \left[ \be(\om) \; z -
\om t \right] \ ,
$\hfill} (23)

\

where coefficients $K_n$ {\em coincide} with the $A_n$ given by
eq.(11) or eq.(15') in the TM ot TE cases, respectively; and
functions $R_n$ are again given by eq.(8) or eq.(13'), respectively;
since the values $k_n$,

\

\hfill{$ \dis{ k_n \equiv \frac{\om^2}{c^2} - \be^2} \ ,
$\hfill} (24)

\

are equal to those found via the (two-dimensional) eq.(9) in the TM and via
eq.(14) in the TE case, simply {\em multiplied} by $\sin\th$ [because of the
fact that, when going on from the two-dimensional membrane to the
three-dimensional coaxial cable, the internal and external radia are equal
(no longer to $a$, $b$, but) to $r_1 = a / \sin \th$ and
$r_2 = b / \sin \th$].

\h Let us now consider the spectral functions

\

\hfill{$
W_n \equiv \exp [-q^2(\om-\om_n)^2] \ ,
$\hfill} (25)

\

with the same weight-parameter $q$, so that $\Delta \om$ too is the
same [according to our definitions, $\Delta \om = 1/q$]; and with

\

\hfill{$ \om_n \equiv \dis{ {{k_n \,c} \over {\sin\th}} } \ ,
$\hfill} (26)

\

quantity $\sin\th$ having a fixed but otherwise arbitrary value. \
We shall construct FTESs, \ ${{\cal F}}(\rho,z;t)$, \ of the
type\footnote{When integrating over $\om$ from $-\infi$ to
$+\infi$ there are also the non-physical (traveling backwards in
space) and the evanescent waves. But their actual contribution is
totally negligible, since the weight-functions $W_n$ are strongly
localized in the vicinity of the $\om_n$-values (which are all
positive: see, e.g., Fig.9). In any case, one could integrate
from $0$ to $\infi$ at the price of incresing a little the
mathematical complexity: we are preferring the present formalism
for simplicity's sake.}

\

\hfill{$
{\cal F}_{3\Drm}(\rho,z;t) \ug \sum_{n=1}^N \; \int_{-\infi}^\infi \,
\drm\om \,
\psi_{n} \, W_n \ ,
$\hfill} (27)

\

with arbitrary $N$. \ Notice that we are not using a single gaussian weight,
but a different gaussian function for each $\om_n$-value, such weights being
centered around the corresponding $\om_n$.

\h Due to the mentioned localization of the $W_n$ around the $\om_n$-values,
we can (for each value of $n$ in the above sum) expand the function
$\be(\om)$ in the neighbourhood of the corresponding $\om_n$-value:

\

\hfill{$
\be(\om) \; \simeq \; \be_{0n} \, + \, \dis{ \left. {{\pa\be} \over
{\pa\om}}
\right|_{\om_n} } (\om - \om_n) + \ ...
$\hfill} (28)

\

where $\be_{0n} = \om_n \cos\th$, and the further terms are neglected
since $\Delta \om$ is assumed to be small. Notice that, because of
relations (26) and (24), in eq.(28) the group-velocities, given by

$$ \dis{ {1 \over {v_{\grm n}}} \ugg  \left. {{\pa\be} \over {\pa\om}}
\right|_{\om_n} } \ , $$

are actually independent of $n$, all of them possessing therefore the same
value:

\

\hfill{$ \dis{ v_{\grm n} \, \equiv \, v_\grm \ug {c\,\cos\th} } \ .
$\hfill} (28')

\

By using relation (28) and the transformation of variables

$$f_n \equiv \om - \om_n \ ,$$

the integration in eq.(27) does eventually yield:

\

\hfill{$
{\cal F}_{3\Drm}(\rho,z;t) \ug \dis{ {\pisr \over q} \exp
\left[ -{{(z-v_gt)^2}
\over {4q^2v_g^2}} \right] } \ \sum_{n=1}^{\infty} A_n \; R_n(\rho) \;
\cos \left[ k_n (z-Vt) \cos\th \right] \ ,
$\hfill} (29)

\

where, let us recall, $V = c^2/v_\grm = c/\cos\th$, and we used the identity

$$
\int_{-\infi}^{\infty} \drm f \; \exp[-q^2 f^2] \; \cos[f (v_g^{-1}z-t)]
\ugg
\dis{ {\pisr \over q} \exp \left[ -{{(v_g^{-1}z-t)^2} \over {4q^2}}
\right] } \ .
$$

\h It is rather interesting to notice that the FTES (29) is related to the
X-shaped waves, since the integration in eq.(27) does eventually yield
the FTES in the form:

\

\hfill{$
{\cal F}_{3\Drm}(\rho,z;t) \ugg \dis{ {\pisr \over q} \exp
\left[ -{{(z-v_gt)^2}
\over {4q^2v_g^2}} \right] } \ {{\cal T}}(\rho,z) \ ,
$\hfill} (30)

\

function ${\cal T}(\rho,z)$ being one of our previous solutions in
eq.(12) or (16) above, at our free choice.

\h Let us go back to the important relation (28'), and to the
discussion about it started at the beginning of this Section. Let us
repeat that, if we choose the $\om_n$-values as in Fig.9, all our
small-bandwidth packets, centered at the $\om_n$'s, get the
same phase-velocity $V>c$ and therefore the same group-velocity
$v_\grm < c$ [since for metallic waveguides the quantities
$k_n^2=\om_n^2/c^2\,-\be^2$ are constant for each mode, and
$v_\grm \equiv \pa\om / \pa\be$, so that it is $V v_\grm=c^2$].
This means that the envelope of solution (29)-(30) moves with
slower-than-light speed; the envelope length\footnote{One may
call ``envelope length" the distance between the two points in
which the envelope height is, for instance, 10\% of its maximum
height.} $\Delta l$ depending on the chosen $\Delta \om$, and
being therefore proportional to $q v_\grm$.

\h However, inside such an envelope, one gets a train of (X-shaped) pulses
---having nothing to do with the ordinary carrier
wave,\footnote{Actually, they can be regarded as a sum of carrier
waves.} as we already mentioned--- traveling with the
Superluminal speed $V$.  An interesting point is that we can
choose the envelope length so that it contains {\em only one}
(X-shaped wave) peak: the Superluminal speed $V=c^2/v_\grm$ of
such a pulse can then be regarded as the actual velocity of
the wave. In order to have just one peak inside the envelope, the
envelope length is to be chosen smaller than the distance between
two successive peaks of the (infinite total energy) train (12),
or (16).

\h It should be noted, at last, that the amplitude of such a
single X-shaped pulse (which remains confined inside the envelope)
first increases, and afterwards decreases, while
traveling; till when it practically disappears. While the
considered pulse tends to vanish on the right (i.e., under the
right tail of the envelope), a second pulse starts to be created
on the left; and so on. \ From eq.(30) it is clear, in fact, that
our finite-energy solution is nothing but an (infinite-energy)
solution of the type in eq.(12), or in eq.(16), multiplied by a
Gaussian function. \ In Figs.10 all such a behaviour is clearly
depicted.

\

\

\centerline{\bf {Acknowledgements}}

\

The authors acknowledge, first of all, very useful discussions with
F.Fontana. \ For stimulating discussions, thanks are due also to V.Abate,
C.Becchi, M.Brambilla, C.Cocca, R.Collina, G.C.Costa, P.Cotta-Ramusino,
C.Dartora, G.Degli Antoni, A.C.G.Fern\'{a}ndez, L.C.Kretly, J.M.Madureira,
G.Pedrazzini, G.Salesi, J.W.Swart, M.T.Vasconselos, M.Villa,
S.Zamboni-Rached and particularly A.Shaarawi.  At last, an anonymous Referee
should be thanked for useful comments.

%%\newpage

\centerline{\bf {Appendix}}

\

\h Let us here state, and demonstrate, the Lu's theorem, for the reader's
convenience:

\h {\bf The theorem\/}: \ Be $\psi_{2\Drm}(x,y;t)$ a solution of the two-dimensional
homogeneous wave equation

\

\hfill{$ \left( \pa_x^2 + \pa_y^2 - \dis{{1 \over {c^2}}} \pa_t^2
\right) \; \psi_{2\Drm}(x,y;t) \ = \ 0 \ . $\hfill} (A.1)

\

\h On applying the transformations

\

\hfill{$ x \lra x' \, \sin \th \ ; \ \ \ y \lra y' \, \sin \th   \
\ \ \et \ \ \ t \lra t' - z' \, \dis{{{\cos \th} \over c}} \ ,
$\hfill} (A.2)

\

the angle $\th$ being fixed ($0 < \th < \pi/2$), the three-dimensional
function

\

\hfill{$ \psi_{3\Drm}(x',y',z';t') =
\psi_{2\Drm}(x'\,\sin\theta\,,\,y'\,\sin\theta\,;\,t'-\cos\theta\,z'/c) 
$\hfill} (A.3)

\

results to be a solution of the three-dimensional wave equation

\

\hfill{$ \left( \pa_{x'}^2 + \pa_{y'}^2 + \pa_{z'}^2 - \dis{{1
\over {c^2}}} \pa_{t'}^2 \right) \; \psi_{3\Drm}(x',y',z'; t') \ = \
0 \ . $\hfill} (A.4)

\

\h {\bf Its demonstration\/}: \ By use of eqs.(A.2), (A.3) and of 
assumption (A.1), one obtains, by direct calculations, that

\

\hfill{$ \begin{array}{l} \left( \pa_{x'}^2 + \pa_{y'}^2 + \pa_{z'}^2 -
\dis{{1 \over {c^2}}} \pa_{t'}^2
\right) \; \psi_{3\Drm}(x',y',z'; t') \ = \\
\\
\left( \sin^2\theta\,\pa_x^2 + \sin^2\theta\,\pa_y^2 + \dis{{\cos^2\theta
\over {c^2}}} \pa_t^2 - \dis{{ 1 \over {c^2}}} \pa_t^2
\right) \; \psi_{2\Drm}(x,y;t) \ =  \\
\\
\sin^2\theta \left( \pa_x^2 + \pa_y^2 - \dis{{1 \over {c^2}}} \pa_t^2
\right) \; \psi_{2\Drm}(x,y;t) \ = \ 0 \ , \end{array} $\hfill}

\

so that the theorem gets demonstrated.

\newpage

\centerline{\bf Figure Captions}

\vspace{1.5cm}

Fig.1 --- Sketch of the coaxial waveguide.

\vspace{1cm}

Fig.2 --- Square magnitude $|\Psi_{2\Drm}(\rho;t=0)|^2$ of the
two-dimensional solutions in eq.(10) for fixed time $(t=0)$ and
for $N=10$ (dotted line) or $N=40$ (solid line). It refers to the
TM case (Dirichlet boundary conditions) with $a=1\;$cm, $b=3\;$cm
and $\rho_0=2\;$cm: See the text.

\vspace{1cm}

Figs.3 --- In Figs.(a) and (b) we show the square magnitude
$|\Psi_{3\Drm}(\rho,\eta)|^2$ of a single (X-shaped)
three-dimensional pulse of the beam in eq.(12), with $\theta =
84^{\orm}$, \ $r_1=a/\sin\theta$, \ $r_2=b/\sin\theta$ (it having been chosen
$a=1\;$cm and $b=3\;$cm), for $N=10$ and $N=40$, respectively. They refer to
the TM case. Notice that $\eta \equiv z-Vt$, and that the
considered beam is a train of X-shaped pulses.

\vspace{1cm}

Fig.4 --- In this figure we depict, by contrast, a {\em couple}
of elements of the train of X-shaped pulses represented in the TM
case by eq.(12), for $N=40$.  This time the angle $\th = 45^\orm$
was chosen, keeping the same $a$ and $b$ values as before.

\vspace{1cm}

Fig.5 --- The orthogonal projection is shown of a single pulse (of the
solution in Fig.4, referring to the TM case) onto the ($\rho,z$) plane
for $t=0$, with $\th = 45^\orm$ and $N=40$.

\vspace{1cm}

Fig.6 --- In analogy with Fig.2, the square magnitude
$|\Psi_{2\Drm}(\rho;t=0)|^2$ is shown of the two-dimensional solutions in
eq.(15) for fixed time $(t=0)$, and for $a=1\;$cm, $b=3\;$cm,
$\rho_0=2\;$cm; this time it refers, however, to the TE case (Neumann
boundary conditions): See the text. \ Again, the dotted line corresponds to
$N=10$, and the solid line to $N=40$.

\vspace{1cm}

Fig.7 --- In this figure, which refers to the TE case, two elements
are depicted of the train of X-shaped pulses represented by
eq.(16), with $\th = 60^\orm$ and $N=40$, while keeping the same $a$ and
$b$ values as before.

\vspace{1cm}

Fig.8 --- The orthogonal projection is shown of a single pulse (of the
solution in Fig.7, for the TE case) onto the plane ($\rho,z$) for $t=0$,
with $\th = 60^\orm$ and $N=40$.

\vspace{1cm}

Fig.9 --- Dispersion curves for the symmetrical $\TM_{0n}$ modes in a
perfect coaxial waveguide, and location of the frequencies whose
corresponding modes possess the same phase-velocity. \ [Actually, the
phase-velocity $c / \cos\th$ of all the terms in expansion (12) is always
the same, being independent of the mode index~$n$: In such a case, it is known
that the group-velocity of the pulse (namely, the velocity {\em tout court}
of the localized pulse) becomes equal to the phase-velocity.]

\vspace{1cm}

Figs.10 --- Time evolution of a {\em finite} total energy
solution. Choosing $q=0.606$ s, $c=1$, $N=40$, $a=1\,$cm,
$b=3\,$cm and $\theta=45^{\orm}$, there is only {\em one} X-shape
pulse inside the subluminal envelope: see the text. The pulse and
envelope velocities are given by $V=1/cos\theta$ and $v_{g}=1/V$:
The superluminal speed $V=1/v_\grm$ of such a pulse can be
regarded, of course, as the actual velocity of the wave. \
Figures (a), (b), (c), (d), (e) and (f) show a complete cycle of
the pulse; they correspond to the time instants $t=0$, $t=0.5
\,$s, $t=1 \,$s, $t=3 \,$s, $t=3.5 \,$s, and $t=4 \,$s,
respectively.

\newpage

REFERENCES\\

[1] M.Zamboni Rached, E.Recami and F.Fontana: ``Localized
Superluminal solutions to Maxwell equations propagating along
a normal-sized waveguide" [Lanl Archives \# physics/0001039],
Phys. Rev. E64 (2001) 066603.\hfill\break

[2] H.Bateman: {\em Electrical and Optical Wave Motion}
(Cambridge Univ.Press; Cambridge, 1915), p.315. \ See also:
J.A.Stratton:  {\em Electromagnetic Theory}  (McGraw-Hill;
New York, 1941), p.356.\hfill\break

[3] A.O.Barut et al.: Phys. Lett. A143 (1990) 349; Found.
Phys. Lett. 3 (1990) 303; Found. Phys. 22 (1992) 1267.\hfill\break

[4] A.O.Barut et al.: Phys. Lett. A180 (1993) 5; A189 (1994)
277.\hfill\break

[5] R.Donnelly and R.W.Ziolkowski: Proc. Roy. Soc. London A440
(1993) 541; \ I.M.Besieris, A.M.Shaarawi and R.W.Ziolkowski:
J. Math. Phys. 30 (1989) 1254; \ S.Esposito: Phys. Lett. A225
(1997) 203; \ J.Vaz and W.A.Rodrigues: Adv. Appl. Cliff. Alg.
S-7 (1997) 457.\hfill\break

[6] R.Courant and D.Hilbert: {\em Methods of Mathematical
Physics} (J.Wiley; New York, 1966), vol.2, p.760. \ Cf. also:
J.N.Brittingham: J. Appl. Phys. 54 (1983) 1179; \
R.W.Ziolkowski: J. Math. Phys. 26 (1985) 861; \ J.Durnin, J.J.Miceli and
J.H.Eberly: Phys. Rev. Lett. 58 (1987) 1499; Opt. Lett. 13 (1988) 79; \ 
A.M.Shaarawi, I.M.Besieris and R.W.Ziolkowski: J. Math. Phys. 31 (1990) 2511; \
P.Hillion: Acta Applicandae Matematicae 30 (1993) 35.\hfill\break

[7] A.O.Barut, G. D.Maccarrone and E.Recami: Nuovo Cimento
A71 (1982) 509; \ E.Recami et al.: Lett. Nuovo Cim. 28 (1980)
151; 29 (1980) 241.\hfill\break

[8] E.Recami: Rivista N. Cim. 9(6) (1986) 1--178. \ Cf. also
E.Recami, F.Fontana and R.Garavaglia: Int. J. Mod. Phys. A15
(2000) 2793; \ E.Recami et al.: Il Nuovo Saggiatore 2(3) (1986)
20; 17(1-2) (2001) 21; \ and Found. Phys. 31 (2001) 1119.\hfill\break

[9] J.-y.Lu and J.F.Greenleaf: IEEE Trans. Ultrason.
Ferroelectr. Freq. Control 39 (1992) 19.\hfill\break

[10] R.W.Ziolkowski, I.M.Besieris and A.M.Shaarawi: J. Opt.
Soc. Am., A10 (1993) 75.\hfill\break

[11] E.Recami: Physica A252 (1998) 586. \ See also J.-y.Lu,
J.F.Greenleaf and E.Recami: ``Limited diffraction solutions
to Maxwell (and Schroedinger) equations'', Lanl Archives e-print
physics/9610012 (Oct.1996). \ Cf. also E.Recami, in {\em Time's Arrows,
Quantum Measurement and Superluminal Behaviour}, ed. by D.Mugnai,
A.Ranfagni and L.S.Shulman (C.N.R.; Rome, 2001), pp.17-36.\hfill\break

[12] M.Zamboni Rached, E.Recami and
H.E.Hern\'aqndez-Figueroa: ``New localized Superluminal solutions to the
wave equations with finite total energies and arbitrary frequencies",
Lanl Archives e-print physics/0109062, to appear in Europ. Phys. Journal
D.\hfill\break

[13] J.-y.Lu and J.F.Greenleaf: IEEE Trans. Ultrason.
Ferroelectr. Freq. Control 39 (1992) 441 \ [in this case the
beam speed is larger than the {\em sound} speed in the
considered medium].\hfill\break

[14] P.Saari and K.Reivelt: Phys. Rev. Lett. 79 (1997)
4135.\hfill\break

[15] D.Mugnai, A.Ranfagni and R.Ruggeri: Phys. Rev. Lett. 84
(2000) 4830. \ [For a panoramic review of the ``Superluminal"
experiments, see E.Recami: Lanl Archives e-print physics/0101108,
Found. Phys. 31 (2001) 1119].\hfill\break

[16] Cf. also, e.g., A.M.Shaarawi and I.M.Besieris, J. Phys.
A: Math.Gen. 33 (2000) 7227; 33 (2000) 7255; 33 (2000) 8559;
Phys. Rev. E62 (2000) 7415.\hfill\break

[17] See, e.g., R.Collins: {\em Field Theory of Guided Waves}
(1991).\hfill\break

[18] J.-y.Lu, H.-h.Zou and J.F.Greenleaf: IEEE Transactions
on Ultrasonics, Ferroelectrics and Frequency Control 42 (1995)
850.\hfill\break

[19] See, e.g., E.Butkov: {\em Mathematical Physics}
(Addison-Wesley; 1968).\hfill\break

[20] See, e.g., J.D.Jackson:  {\em Classical Electrodynamics}
(J.Wiley; New York, 1975).\hfill\break

[21] Cf., e.g., S.Ramo, J.R.Whinnery and T.Van Duzer: {\em Fields and Waves
in Communication Electronics}, Chapt. 8 (John Wiley; New York,
1984).\hfill\break

[22] Cf., e.g., ref.[11] and refs. therein.\hfill\break

[23] See, e.g., M.Zamboni Rached and H.E.Hern\'andez-Figueroa: Optics Comm.
191 (2000) 49. \ From the experimental point of view, cf., e.g., S.Longhi,
P.Laporta, M.Belmonte and E.Recami: ``Measurement of superluminal optical
tunnelling in double-barrier photonic bandgaps", Phys. Rev. E65 (2002) 046610. \
Cf. also V.S.Olkhovsky, E.Recami and G.Salesi: ``Tunneling through two successive
barriers and the Hartman (Superluminal) effect", Europhys. Lett. 57 (2002)
879-884; \ Y.Aharonov, N.Erez and B.Reznik: ``Superoscillations and tunnelling
times", Phys. Rev. A65 (2002) 052124.\hfill\break

[24] I.M.Besieris, M.Abdel-Rahman, A.Shaarawi and A.Chatzipetros:
Progress in Electromagnetic Research (PIER) 19 (1998) 1-48
(1998).\hfill\break

\end{document}